# Sign reversal of Berry curvature triple driven by magnetic phase transition in a ferromagnetic polar metal

Xuyang Sha[1,2,†], Xuejin Zhang[3†], Hao Liu[1,2,†], Jin Cao[3†], Ruohan Chen[2], Jinfeng Zhai[1,2], Dingfu Shao[4], Shiwei Wu[1,2], Cong Xiao[5*], Shengyuan A. Yang[3*], Pan He[1,6,7*], Hangwen Guo[1,6,7*] and Jian Shen[1,2,6,7,8*]

[1] *State Key Laboratory of Surface Physics and Institute for Nanoelectronic Devices and Quantum Computing, Fudan University, Shanghai 200433, China*

[2] *Department of Physics, Fudan University, Shanghai 200433, China*

[3] *Research Laboratory for Quantum Materials, IAPME, FST, University of Macau, Macau SAR, China*

[4] *Key Laboratory of Materials Physics, Institute of Solid State Physics, HFIPS, Chinese Academy of Sciences, Hefei 230031*

[5] *Interdisciplinary Center for Theoretical Physics and Information Sciences (ICTPIS), Fudan University, Shanghai 200433, China*

[6] *Shanghai Research Center for Quantum Sciences, Shanghai 201315, China*

[7] *Zhangjiang Fudan International Innovation Center, Fudan University, Shanghai 201210, China*

[8] *Collaborative Innovation Center of Advanced Microstructures, Nanjing 210093, China*

## Abstract

**While the time-reversal-even ($\mathcal{T}$-even) nonlinear Hall effect has been extensively discussed in nonmagnetic materials, the impact of magnetic phase transition on it remains largely overlooked. Here, we report an abrupt enhancement of the $\mathcal{T}$-even nonlinear Hall effect in non-centrosymmetric $SrRuO_3$(111) thin films during the paramagnetic-ferromagnetic transition. Scaling analysis reveals a sign reversal of both the skew scattering and side jump contributions upon the emergence of magnetism, which we attribute to the sign change of Berry curvature triple on the Fermi surface. Density functional theory calculations support this interpretation, ascribing this behavior to distinctive origins of Berry curvature hot spots in paramagnetic and ferromagnetic phases. Our findings unveil the exchange-induced dramatic nonperturbative change of nonlinear Hall effect, and establish $SrRuO_3$(111) thin films as a promising platform for exploring magnetically tunable nonlinear transport effects.**

The interplay between magnetism and Berry curvature is a key factor in understanding exotic transport phenomena in magnetic systems [1,2]. In linear response, the magnetic phase transition lifts spin degeneracy and enables generation of Berry curvature monopoles in time-reversal breaking band structures, leading to anomalous Hall effect [3–7] and anomalous thermoelectric transport [8]. In contrast to linear Hall responses, nonlinear Hall effect (NHE) is allowed in nonmagnetic systems without magnetic field, which is termed as time-reversal-even ($\mathcal{T}$-even) response, and has attracted significant attention in recent years [9–13]. While inversion symmetry breaking is known to be a prerequisite for $\mathcal{T}$-even NHE, the impact of broken time-reversal symmetry on this effect has remained largely unnoticed. Remarkably, NHE is known to be sensitive to crystalline order and can changes dramatically during topological or ferroelectric phase transitions [14–19]. Therefore, it is conceivable that a magnetic phase transition can also change the behavior of $\mathcal{T}$-even NHE. In particular, if the exchange coupling reshapes the hot-spot structures of Berry curvature around the Fermi surface, then the NHE is anticipated to be fundamentally altered. This appealing possibility has not been explored yet. Once achieved, this will not only illuminate the so far obscure physics of exchange-driven nonlinear Berry curvature effects, but also unlock new potential for harnessing nonlinear responses to create novel spintronics device functionalities [10,20].

Complex oxides offer an ideal platform for exploring this possibility. Their strong charge-spin-lattice coupling enables the engineering of a wide range of structural, electronic, and magnetic properties, and allows flexible control over both inversion and time-reversal symmetries [21–24]. It was shown that strain engineering in ferromagnetic metallic oxides can induce electrical polarity, stabilizing a magnetic polar metal phase [25–28], which creates an ideal platform to investigate nonlinear transport in ferromagnets. Interestingly, in thin films [29,30], a threefold rotational axis can be maintained along the growth direction, which prohibits the dipole form of Berry curvature distribution but enforces a novel Berry curvature triple (BCT) pattern [31–34]. In such a case, the $\mathcal{T}$-even NHE can only arise from disorder induced extrinsic mechanisms such as skew scattering and side jump [35–37], which are intimately linked to BCT [31–34]. As a geometric quantity uniquely characterizing threefold symmetric Berry curvature distributions, BCT remains unexplored in magnetic phases and upon phase transitions. Therefore, examining how magnetic phase transitions affect $\mathcal{T}$-even NHE is of significant interest for elucidating exchange-driven variations in BCT.

In this work, we report the observation of $\mathcal{T}$-even NHE in polar metal $SrRuO_3$ (111) thin films. The $SrRuO_3$(111) films exhibit inversion symmetry breaking by flexoelectric effect [38] but preserve an effective threefold rotational symmetry in the transport plane, allowing for nonvanishing BCT. During the paramagnetic-ferromagnetic phase transition at 138 K, we observed a large enhancement of nonlinear Hall conductivity. Remarkably, the scaling analysis shows that when the system enters into ferromagnetic phase, both the skew scattering and side jump nonlinear Hall contributions reverse their signs, which are well accounted for by our first-principles calculations revealing a sign change of BCT across the magnetic phase transition. The calculations also suggest that this sign reversal can be traced back to the distinct nature of Berry curvature hot spots in the two phases, i.e., spin-orbit local splitting in paramagnetic phase versus exchange splitting in ferromagnetic phase. This qualitative change is rooted in the emergent strong internal magnetization, going beyond the perturbative picture for nonlinear Hall transport induced by external magnetic field [39–41]. Our observations illuminate nonlinear transport behaviors during magnetic phase transitions and underscore the need for further investigations into the influence of magnetism on $\mathcal{T}$-even NHE.

$SrRuO_3(111)$ thin films were grown on $SrTiO_3(111)$ substrate via pulsed laser deposition technique (See methods for details). As shown in Fig. 1(a), standard Hall bar devices were fabricated via photolithography for transport measurements. The first ($V^{1\omega}$) and second order harmonic voltages ($V^{2\omega}$) can be obtained under an ac source current $I_x = I\ sin\ \omega t$ using lock-in techniques. Fig. 1(b) shows the longitudinal resistivity $\rho_{xx}$ as function of temperature for the thin film with 8 nm thickness. The film shows typical metallic behavior as the $\rho_{xx}$ monotonically decreases with lowering temperature. A kink at 138 K indicates the paramagnetic-ferromagnetic phase transition. To further verify the transition, we measured anomalous Hall effect at different temperatures. We plot the Hall resistivity $\rho_{xy}$ as function of out-of-plane magnetic field for several representative temperatures in Fig. 1(c). Above 138 K, the anomalous Hall effect is negligible indicating the absence of magnetic order. Below 138 K, the anomalous Hall signal start to emerge and exhibit hysteresis behavior, consistent with the ferromagnetic metal properties reported in $SrRuO_3$ thin films [7,42]. $\rho_{xy}$ has a very small contribution from the contact misalignment of the Hall bar devices

(Supplementary Fig. S1).

**Fig. 1. (a)** Schematic of hall-bar device of $SrRuO_3$ thin film. A sinusoidal current $I^{\omega}$ was applied, and the first harmonic $V^{1\omega}$ and second harmonic $V^{2\omega}$ voltages were simultaneously measured along the longitudinal and transverse directions. **(b)** The longitudinal resistivity $\rho_{xx}$ as a function of temperature. **(c)** Magnetic field-dependent

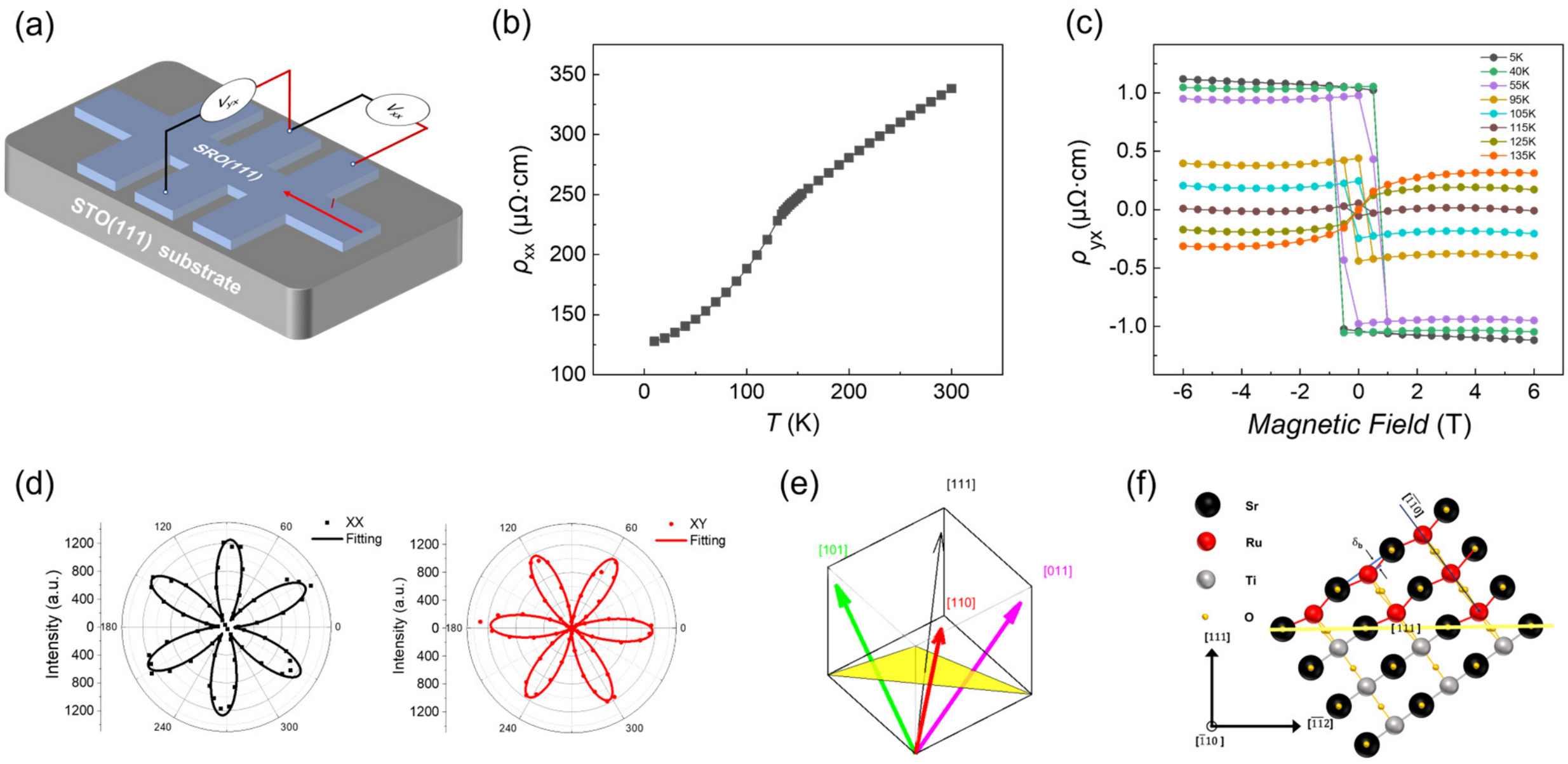

linear Hall resistivities ($\rho_{yx} \sim H$) of 8 nm $SrRuO_3$ thin film. **(d)** SHG patterns acquired under XX and XY polarization configurations. Here, the excitation and detection beams were linearly polarized, with XX and XY referring to co- and cross-linearly polarized between the two beams, respectively. **(e)** Schematic of polar symmetry in $SrRuO_3$ (111) thin film with three electric dipoles along [101], [011] and [110] directions. **(f)** Schematic showing the off-centering of Ru atoms along $[\bar{1}\bar{1}0]$ induced by a flexoelectric field [38].

Strain gradient due to heteroepitaxy can induce non-centrosymmetric structure that breaks inversion symmetry [43]. In $SrRuO_3$ (111) thin film, recent report shows that shear strain gradient causes large Ru ion off-centering displacement along $[\bar{1}\bar{1}0]$ direction, as illustrated in Figs. 1(e) and 1(f) [38]. Such displacement gives rise to a non-centrosymmetric polar structure. To examine its existence, we performed optical second-harmonic generation (SHG) and the results are shown in Fig. 1(d). Threefold symmetric SHG anisotropy pattern is observed which aligns well with point group $m$ with three equivalent electric dipoles originated from 120° rotated domains as illustrated in Fig. 1(e), and is in good agreement with previous report [44]. These results confirm the emergence of inversion-symmetry breaking in a ferromagnetic metallic system, offering a rare breed to explore $\mathcal{T}$-even NHE under time-reversal symmetry breaking conditions.

Fig. 2(a) shows the second-harmonic transverse voltage $V_y^{2\omega}$ measured at zero magnetic field and room temperature. $V_y^{2\omega}$ shows quadratic current dependence $V_y^{2\omega} \propto I^2$, and changes sign when reversing simultaneously the current and corresponding Hall probe directions, validating the nonlinear origin. We further verified that $V_y^{2\omega}$ signal is independent of the ac frequency used (Supplementary Fig. S2), suggesting negligible capacitive contributions. These results unambiguously demonstrate the existence of transverse nonlinear transport in $SrRuO_3$ (111) thin film and are in line with its inversion-symmetry breaking nature [38,44].

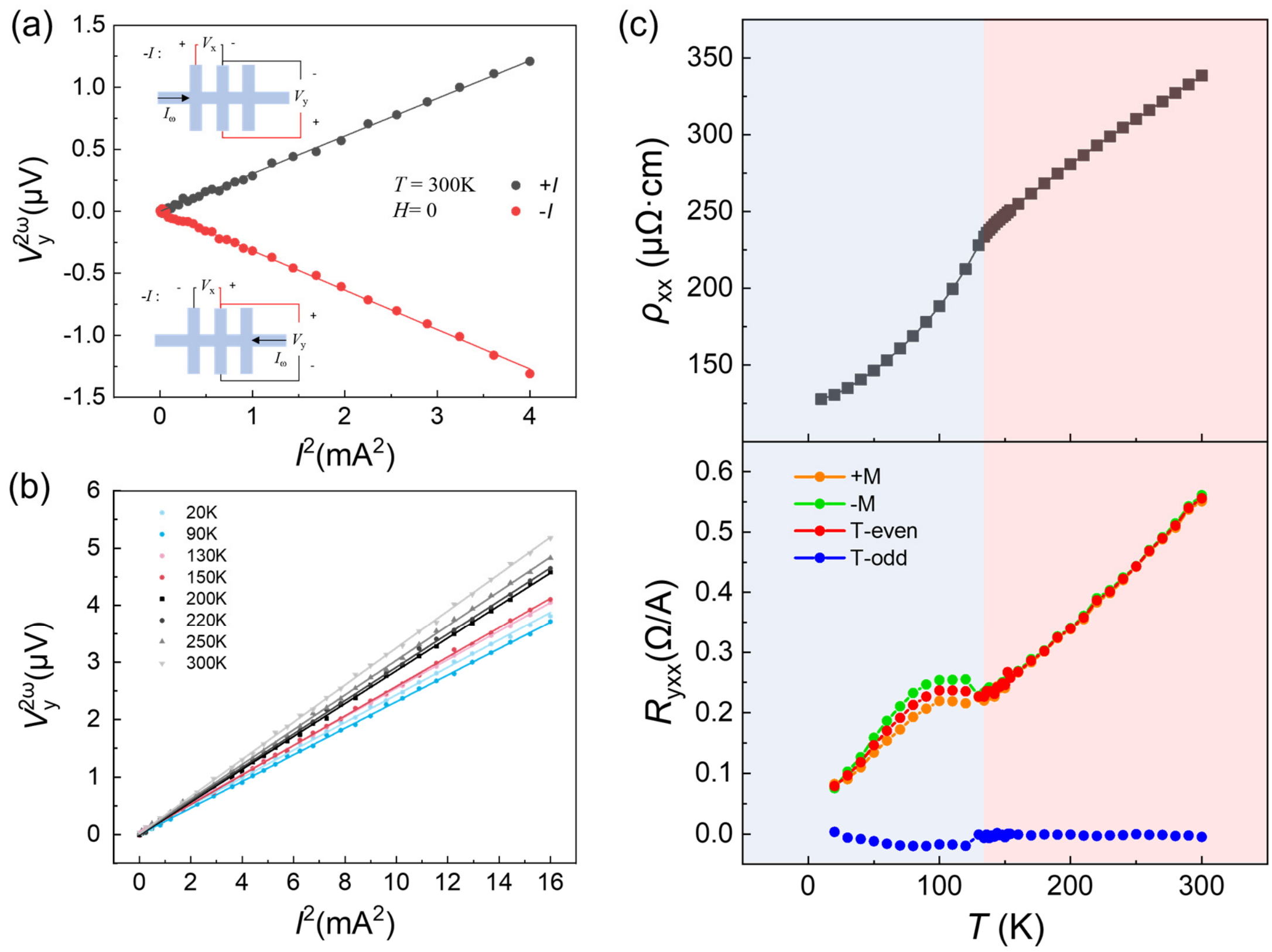

**Fig. 2.** (a) The second-harmonic transverse voltage $V_y^{2\omega}$ versus $I^2$ for the current along the [-1-12] axis. The solid lines are linear fits to the data. The insets show schematics of current and voltage probe connections under opposite current directions of $+I$ and $-I$. The data were collected at $T = 300$ K without external magnetic field. (b) $V_y^{2\omega}$ versus $I^2$ at different temperatures, and the slope of which being the nonlinear transverse resistance $R_{yxx}$. (c) The longitudinal resistivity $\rho_{xx}$ (black), the nonlinear transverse resistance $R_{yxx}$ for upward magnetization (orange) and downward magnetization (green), the T-even term (red) and the T-odd term (blue) of $R_{yxx}$ versus temperature plotted along. Red and blue regions denote the high temperature paramagnetic and low temperature ferromagnetic phase, respectively.

We then investigated the temperature dependence of nonlinear transport. Figure 2(b) shows the $V_y^{2\omega} \sim I^2$ curves at different temperatures under zero magnetic field, all of which show good quadratic current dependence. The slope of the $V_y^{2\omega} \sim I^2$ curves is defined as the second order Hall resistance $R_{yxx}$, which is plotted along with $\rho_{xx}$ in Fig. 2(c). We achieve $R_{yxx}$ for both upward and downward magnetization ($M$) at zero magnetic field. The $\mathcal{T}$-even term is defined as $R_{yxx}^{T-even} = \frac{R_{yxx}(+M)+R_{yxx}(-M)}{2}$ while the $\mathcal{T}$-odd term is defined as $R_{yxx}^{T-odd} = \frac{R_{yxx}(+M)-R_{yxx}(-M)}{2}$. While $R_{yxx}^{T-even}$ can exist in both paramagnetic and ferromagnetic phases, $R_{yxx}^{T-odd}$, which may arise from thermally driven anomalous Nernst effect or nonlinear $\mathcal{T}$-odd NHE [45–47], is only allowed in ferromagnetic phase. As shown in Fig. 2(c), $R_{yxx}^{T-odd}$ is indeed vanishingly small in the paramagnetic phase, and remarkably, is still negligible compared to $R_{yxx}^{T-even}$ in the ferromagnetic phase, showing the dominant $\mathcal{T}$-even contribution to NHE. The smallness of $R_{yxx}^{T-odd}$ in ferromagnetic phase is found in all the measured devices (Supplementary Fig. S3), thus solidly validating the $\mathcal{T}$-even nature of the observed nonlinear transport in both paramagnetic and ferromagnetic phases. Therefore, $R_{yxx} \simeq R_{yxx}^{T-even}$ and in the following analysis we use the $R_{yxx}^{T-even}$ data for $R_{yxx}$.

To gain more insight into the impact of paramagnetic-ferromagnetic phase transition, we evaluated the nonlinear transverse conductivity $\sigma_{yxx}^{(2)} = \sigma_{xx} \frac{V_y^{2\omega}}{(V_x^{\omega})^2} \frac{L^2}{W}$, where $L$ and $W$ are the length and width of the Hall bar channel. The results are shown in Fig. 3(a). While $\sigma_{yxx}^{(2)}$ varies slowly above the magnetic transition temperature $T_c$, a sharp increase with decreasing temperature is clearly observed in the ferromagnetic phase. We then analyzed the scaling behavior to help identify the involved physical mechanisms [11,31,35]. Using the temperature-dependent data, $\sigma_{yxx}^{(2)}/\sigma_{xx}$ is plotted as a function of $\sigma_{xx}^2$ as shown in Fig. 3(b). Two regimes are clearly observed corresponding to the paramagnetic ( $\sigma_{xx}^2 < 16\times10^6$ (S/cm)$^2$) and ferromagnetic ($\sigma_{xx}^2 > 16\times10^6$ (S/cm)$^2$) phases from the transport data. In both regimes we found that the data well satisfy the scaling

$$\sigma_{yxx}^{(2)} = \xi\sigma_{xx}^3 + \eta\sigma_{xx}. \quad (1)$$

Here, the cubic term corresponds to the contribution from skew scattering [10,15], whereas the linear term may originate from Berry curvature dipole [9,11] or side jump [11,31]. Considering that the system has a threefold rotational symmetry evidenced by the results in Fig.1, the contribution of Berry curvature dipole is excluded. We extracted the values of $\xi$ and $\eta$ from the fitting. An interesting finding is that both $\xi$ and $\eta$ experience a sign change during the paramagnetic-ferromagnetic transition. Such sign-changing scaling behaviors were cross-checked in all other samples (Supplementary Fig. S4 and Fig. S5). In particular, thick samples (25nm) also display the sign change of both $\xi$ and $\eta$, validating that such behaviors do not originate from surface or interface, but rather reflects intrinsic nonlinear physics of bulk states during the magnetic phase transition. Moreover, in temperature-dependent SHG measurements (Supplementary Fig. S6) [48], we do not observe any structural change or non-collinear magnetic inhomogeneity during the magnetic phase transition, pointing towards an electronic band origin of our observation. Here, we mention that Eq. (1) has been employed by a recent work on NHE in Sb-doped topological insulator $MnBi_4Te_7$ [49], where a sign change of $\xi$ was also claimed. However, it is unknown whether the signal measured there is $\mathcal{T}$-even or $\mathcal{T}$-odd. If it is $\mathcal{T}$-odd, then the cubic term has totally different mechanisms from skew scattering [50]. More critically, the linear longitudinal resistivity therein shows an insulating feature, thus questioning the use of Eq. (1), which does not apply to such case [6,35,51]. On the other hand, in our case the resistivity shows metallic behavior (Fig. 2(c)), justifying the use of Eq. (1).

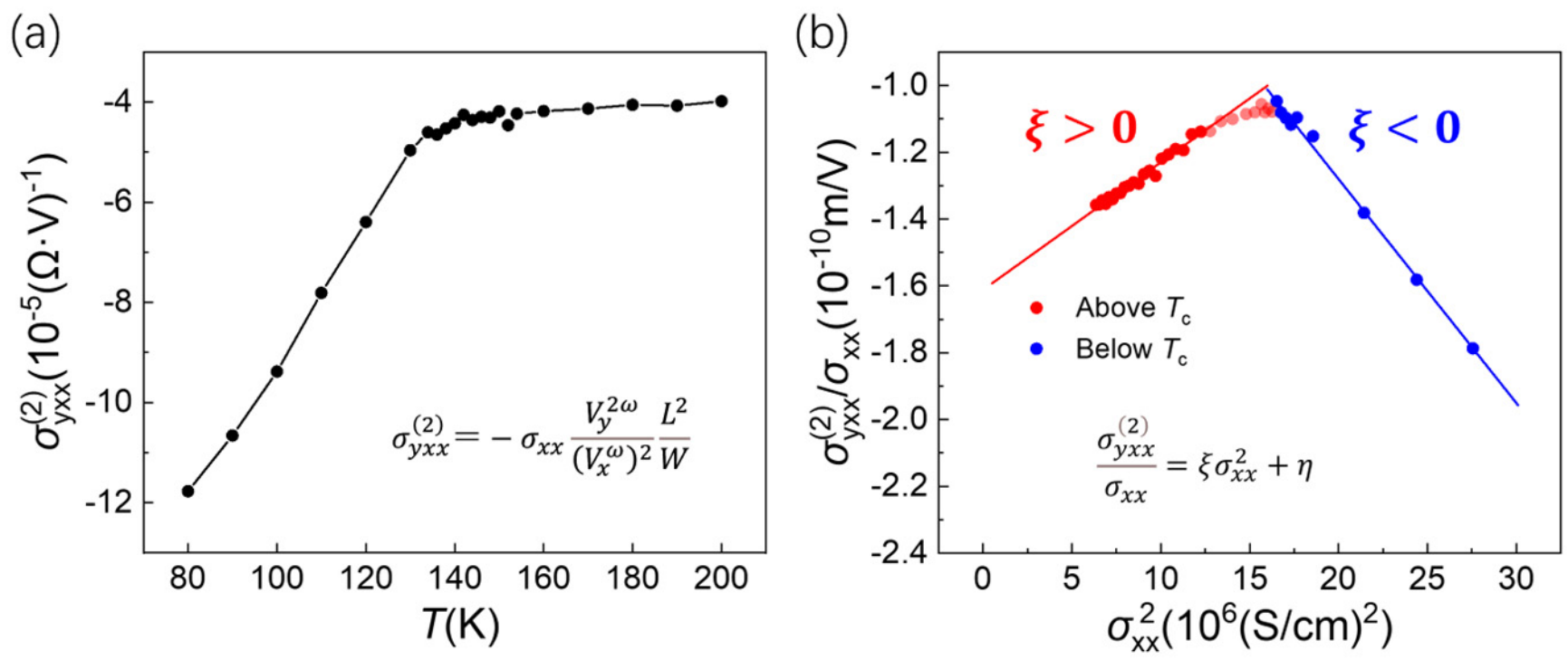

(c)

| | $\xi$ ($10^{-22}$ $m^3\cdot V^{-1}S^{-2}$) | $\eta$ ($10^{-10}$ $m\cdot V^{-1}$) |
|---|---|---|
| Above $T_c$ | 3.8±0.1 | -1.61±0.1 |
| Below $T_c$ | -6.7±0.2 | 0.06±0.03 |

**Fig. 3.** (a) The nonlinear transverse conductivity $\sigma^{(2)}_{yxx}$ as a function of temperature. (b) Scaling law analysis by plotting $\sigma^{(2)}_{yxx}/\sigma_{xx}$ versus $\sigma^2_{xx}$. The dots are the experimental data. The solid lines are linear fits to the experimental data in paramagnetic (red) and ferromagnetic (blue) phases. (c) The slopes $\xi$ and intercepts $\eta$ of linear fitting for paramagnetic (red) and ferromagnetic (blue) phases are shown.

As both the skew scattering and side jump are related to BCT under a threefold rotational symmetry [16-19], the sign change of $\xi$ and $\eta$ can be traced back to that of BCT, which we explored from symmetry characters and first-principles calculations. The BCT is defined as a convolution of Berry curvature $\Omega_z(\boldsymbol{k})$ with an angular form factor characteristic of the threefold rotational symmetry [31–34]: $\mathrm{BCT}(\mu) = 2\pi\hbar\int\frac{d\boldsymbol{k}}{(2\pi)^2}\delta(\varepsilon_{\boldsymbol{k}}-\mu)\Omega_z(\boldsymbol{k})\cos 3\theta_{\boldsymbol{k}}$, where the integration is over the (111) plane and $\theta_{\boldsymbol{k}}$ is the angle between $\boldsymbol{k}$ and the $\Gamma$-$K$ direction. The structure of the $SrRuO_3$ (111) plane is shown in Fig. 4(a). It exhibits inversion symmetry along with threefold rotation and mirror symmetries, as indicated by the dashed lines. The combined inversion and time-reversal ($\mathcal{PT}$) symmetry enforces double degeneracy in the band structure, leading to vanishing Berry curvature and BCT. As mentioned, the $SrTiO_3$ (111) substrate imposes shear strain on $SrRuO_3$, and the Ru atoms undergo an off-center shift along $[\bar{1}\bar{1}0]$, $[\bar{1}0\bar{1}]$ or $[0\bar{1}\bar{1}]$ direction by 0.1 Å on average in polar domains [38]. As a result, the overall system retains $C_{3z}$ symmetry while the inversion symmetry is broken, enabling nonzero BCT. It should be noted that the electrons on Fermi surfaces are mainly derived from Ru atoms. As such, the shift of Ru atoms shall significantly alter the transport properties of $SrRuO_3$ thin films.

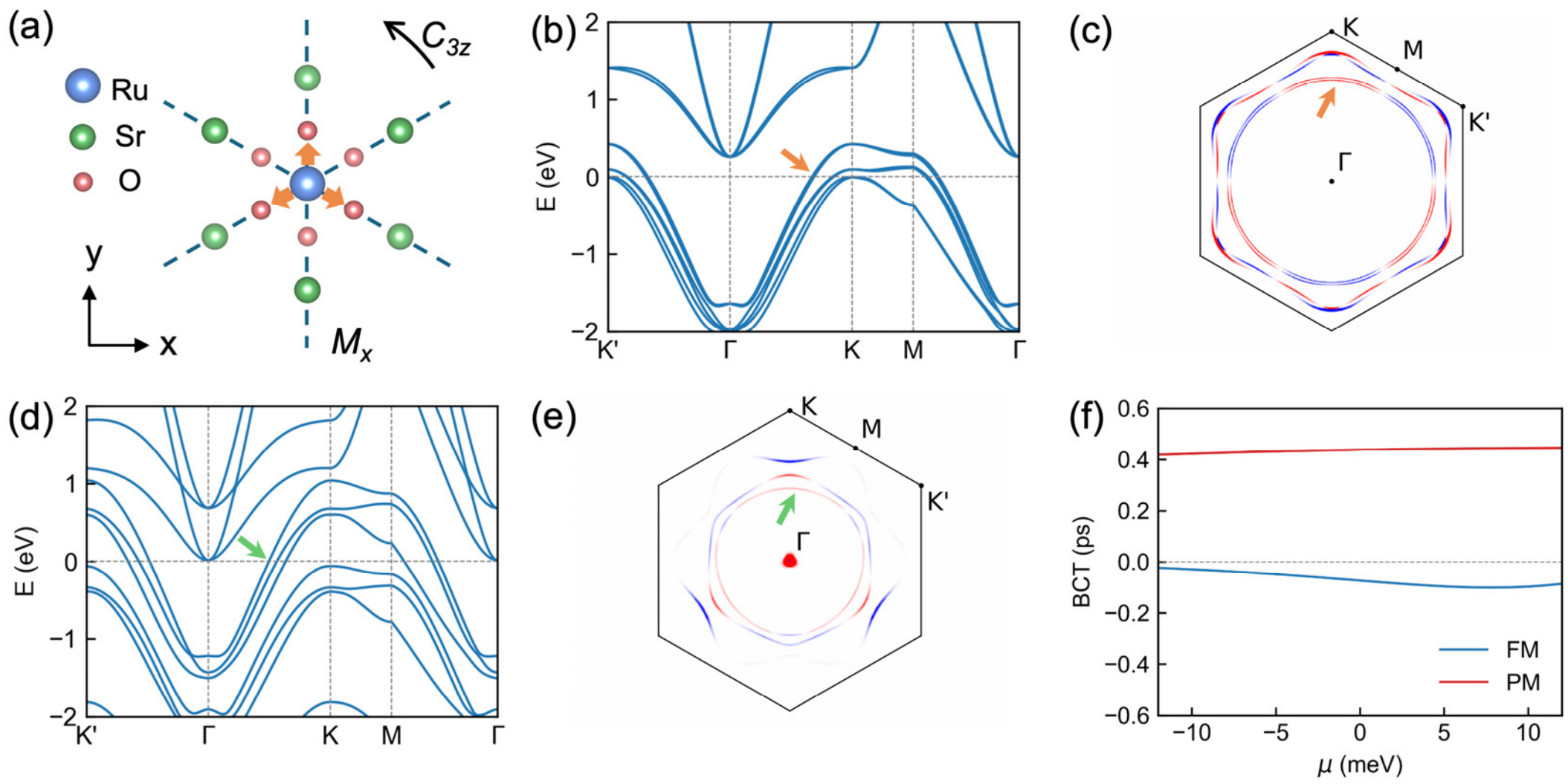

**Fig. 4.** (a) Top view of $SrRuO_3$ (111). The three vectors represent the displacement of Ru atoms along $[\bar{1}\bar{1}0]$, $[\bar{1}0\bar{1}]$ and $[0\bar{1}\bar{1}]$ for the three polar domains. (b), (c) Calculated band structure and Berry curvature distribution at Fermi level for the paramagnetic phase, respectively. (d), (e) Same as (b), (c) but for the ferromagnetic phase. The spin-orbit gap in the paramagnetic phase and exchange gap in the ferromagnetic phase are indicated by orange arrows and green arrows, respectively. (f) Calculated BCTs for paramagnetic (PM) and ferromagnetic (FM) phase as a function of chemical potential.

To perform first-principles calculations, we directly shift the Ru atoms along the [111] direction in the crystal model to simulate the average effect, instead of the realistic displacements as given in Fig. 4(a). The resulting structure breaks inversion symmetry but preserves the threefold axis $C_{3z}$, which is in line with our SHG result. We set the displacement to be 0.1 Å, which was selected based on the previous report [38]. We first examined the paramagnetic phase above the critical temperature. Due to $\mathcal{PT}$ symmetry breaking, there is a band splitting of approximately 40 meV on the Fermi surface, as shown in Fig. 4(b). Such a small splitting originates from the spin-orbit coupling. It enables sizable Berry curvature distributions as shown in Fig. 4(c), which respects $C_{3z}$. In contrast, in the ferromagnetic phase below the critical temperature, the band splitting is significantly larger than the paramagnetic case, as it originates from the exchange interaction associated with Ru atoms. The large band separation is also evident on the Fermi surface, as shown in Fig. 4(e), which reveals a very different pattern of Berry curvature. We then calculated the BCT in both paramagnetic and ferromagnetic phases. As shown in Fig. 4(f), around the chemical potential, the BCT exhibits opposite signs for the two phases. This result is consistent with our observed sign reversal of the skew scattering and side jump contributions across the magnetic phase transition. Moreover, our first-principles results suggest that this sign reversal originates from the distinct mechanisms generating Berry curvature in the two phases (spin-orbit gap versus exchange gap).

In summary, in polar metal $SrRuO_3$(111) film, we reported an abrupt change of $\mathcal{T}$-even nonlinear transverse transport and an unexpected sign reversal driven by a magnetic phase transition. The observations match well with the theoretical picture of exchange driven change of BCT. Our results not only open the door to explore nonlinear physics enabled by BCT in ferromagnetic systems, but also uncover a large number of opportunities to explore the impact of magnetic phase transition on various novel $\mathcal{T}$-even charge [52–55]/spin [56,57] nonlinear transport effects and the engineering of them for oxide-spintronic devices [58].

## Methods

**Sample Fabrication and characterization.** $SrRuO_3$ thin films were grown on $SrTiO_3$(111) substrates using a pulsed-laser deposition system with a KrF excimer laser (248nm). Before deposition, $SrTiO_3$ substrates (miscut <0.1°) were etched with a buffered hydrofluoric acid solution and then annealed in air at 1,050 °C for 1 h to produce an atomically flat surface with a unit-cell step terrace structure. During deposition, the temperature of the substrate was maintained at 650° C and $SrRuO_3$ films were grown under an oxygen pressure of 80 mTorr with a laser fluence of 2 J $cm^{-2}$ and a frequency of 5Hz. A high-pressure reflection high-energy electron diffraction system was used to monitor the growth. A smooth surface is observed as shown in Supplementary Fig. S7, which is further confirmed by Atomic force microscopy (Park NX10 with Al-coated tips). X-ray diffraction Φ scan was performed using a Bruker D8 Discover system.

**Device fabrication.** Conventional photolithography and ion beam etching were used to pattern the $SrRuO_3$ films into the Hall bar geometry. The channel size was minimized to 50×187.5 $\mu m^2$. After patterning, the samples were ex-situ annealed at 300 ° C in ambient oxygen flow for 3 h to minimize the oxygen deficiency induced during growth and ion-milling. Cr (5 nm) and Au (60 nm) films were sputtered onto the Hall bar as contact electrodes. After the entire fabrication process, the SRO conducting channels still exhibit atomically-flat topography (Supplementary Fig. S7).

**Electric transport measurements.** The transport data were measured using a physical properties measurement system (PPMS, Quantum Design). We performed low-frequency ac electric harmonic measurements using Keithley current source (6221) and Stanford Research lock-in amplifiers (SR830). During the measurements, a sinusoidal current with a frequency f = ω/2π = 17.3 Hz is applied to the devices, and the in-phase (0°) first harmonic voltage $V^{\omega}$ and out-of-phase (90°) second harmonic voltage $V^{2\omega}$ along the longitudinal and transverse voltages were measured simultaneously by four lock-in amplifiers.

**Optical second harmonic generation measurements.**

The nonlinear SHG measurements were conducted in a variable-temperature optical cryostat housed inside a superconducting magnet with a room-temperature bore. The excitation beam (120 fs, 80 MHz, 820 nm) from a Ti:sapphire oscillator (MaiTai HP, Spectra Physics) was focused onto the sample at normal incidence using a 50× microscopic objective (NA = 0.55). The reflected nonlinear optical signal was collected by the same objective and detected by either a photomultiplier tube in photon-counting mode or a spectrograph equipped with a liquid-nitrogen-cooled charge coupled device. To measure the polarization-resolved intensity of nonlinear optical signal, the polarization of the excitation and signal beams was rotated with respect to the sample simultaneously using a half-wave plate. The polarization state of the signal beam was analyzed by a linear polarizer placed in front of detectors. For the measurements of temperature dependence and magnetic hysteresis loops, the excitation beam was circularly polarized using a quarter-wave plate.

**Computational details.**

The first-principles calculations were performed using the Vienna Ab initio Simulation Package [59–61]. The projector-augmented wave method [62] was employed to describe the interactions between ions and electrons. Exchange-correlation effects were treated self-consistently using the Perdew-Burke-Ernzerhof functional [63]. The cutoff energy was set to be 500 eV. The Brillouin zone was sampled by a $k$-point grid of $9 \times 9 \times 9$. The convergence threshold for energy and force were set to $10^{-6}$ eV and 0.01 eV/Å, respectively. The spin-orbit coupling was considered in all calculations. The *ab initio* tight-binding model was constructed using the Wannier90 package [64], which was used for calculating the Berry curvature and the BCT.

## Acknowledgement

## Conflict of interest

The authors declare no competing interests.